# Multi-Task Learning for Arousal and Sleep Stage Detection Using Fully Convolutional Networks

Hasan ZAN[a*], Abdulnasır YILDIZ[b]

[a*] Vocational School, Mardin Artuklu University, Mardin, Turkey,
hasanzan@artuklu.edu.tr

[b] Department of Electrical and Electronics Engineering, Dicle University, Diyarbakir, Turkey,
abnayil@dicle.edu.tr

## Abstract

*Objective*. Sleep is a critical physiological process that plays a vital role in maintaining physical and mental health. Accurate detection of arousals and sleep stages is essential for the diagnosis of sleep disorders, as frequent and excessive occurrences of arousals disrupt sleep stage patterns and lead to poor sleep quality, negatively impacting physical and mental health. Polysomnography is a traditional method for arousal and sleep stage detection that is time-consuming and prone to high variability among experts. *Approach*. In this paper, we propose a novel multi-task learning approach for arousal and sleep stage detection using fully convolutional neural networks. Our model, FullSleepNet, accepts a full-night single-channel EEG signal as input and produces segmentation masks for arousal and sleep stage labels. FullSleepNet comprises four modules: a convolutional module to extract local features, a recurrent module to capture long-range dependencies, an attention mechanism to focus on relevant parts of the input, and a segmentation module to output final predictions. *Main results.* By unifying the two interrelated tasks as segmentation problems and employing a multi-task learning approach, FullSleepNet achieves state-of-the-art performance for arousal detection with an area under the precision-recall curve of 0.70 on Sleep Heart Health Study and Multi-Ethnic Study of Atherosclerosis datasets. For sleep stage classification, FullSleepNet obtains comparable performance on both datasets, achieving an accuracy of 0.88 on the former and an accuracy of 0.83 on the latter. *Significance.* Our results demonstrate that FullSleepNet offers improved practicality, efficiency, and accuracy for the detection of arousal and classification of sleep stages using raw EEG signals as input.

# 1. Introduction

Sleep is a fundamental physiological process that plays a crucial role in maintaining physical and mental health [1]. During sleep, the body and brain undergo various changes that support physical and mental restoration [2]. Sleep is characterized by alternating periods of rapid eye movement (REM) sleep and non-rapid eye movement (NREM) sleep, which occur in a repeating pattern throughout the night [3, 4]. A typical sleep stage pattern consists of a progression from lighter stages of NREM sleep to deeper stages of NREM sleep, followed by a period of REM sleep [4]. This pattern typically repeats throughout the night, with the proportion of time spent in each stage varying depending on the individual.

Sleep arousal is a shift from deep to light sleep or from sleep to wakefulness [5]. Arousals can be caused by a variety of factors, including external stimuli such as noise or light or internal factors such as sleep disorders, pain, or discomfort [6]. Frequent and excessive occurrences of arousals, which are considered a sign of sleep disorders [7], can disrupt sleep stage patterns [8, 9]. Since REM and NREM sleep are responsible for mental and physical restoration of the body, respectively, impaired sleep stage patterns lead to poor sleep quality and have negative impacts on physical and mental health [10]. Thus, accurate detection of arousals and sleep stages is of great importance for the diagnosis of sleep disorders.

Traditionally, sleep stage and arousal detection or scoring have been performed using polysomnography (PSG), a method that involves the simultaneous recording of multiple physiological signals, including electroencephalography (EEG), electromyography (EMG), electrocardiogram (ECG), and electrooculography (EOG) [11]. It requires that sleep experts carefully examine and analyze the whole-night PSG signals according to the American Academy of Sleep Medicine (AASM) scoring manual [12]. The manual sets rules for the detection of each stage and arousal event. Stage scoring involves assigning a stage label (one of wakefulness, N1, N2, N3, and REM) to each 30-second-long segment (called an "epoch"), while arousal scoring requires the localization of each arousal event, i.e., determining the onset and duration of each arousal [13]. Although PSG is considered the gold standard for sleep scoring, this process is tedious and time-consuming, and its results demonstrate high variability among even experienced experts due to individual interpretation of the manual [13, 14]. Therefore, the development of computational methods that could improve the reliability and speed of the sleep scoring process has drawn a lot of interest in recent decades.

Many computational approaches have been proposed for automatic sleep scoring. In particular, methods based on manual feature extraction techniques have been extensively employed [13, 15, 16]. These techniques involve features based on time, frequency, time-frequency domain signals, or non-linear parameters [17, 18, 19, 20, 21, 22, 23, 24, 25]. Extracted features have been classified using various algorithms, i.e., k-nearest neighbors [26, 27], support vector machines [25, 28, 29, 30], ensemble classifiers [31, 32, 33], decision trees [27, 34], artificial neural networks [25, 29, 35], and threshold-based classifiers [22, 23, 24, 36]. These earlier works have demonstrated that automatic detection of arousals and sleep stages can be achieved by employing methods relying on handcrafted features. However, because the features have been tailored based on the unique PSG system and available datasets, the associated techniques could need further manual tuning when used with different recording systems [37]. Furthermore, the methods proposed for arousal detection have been developed on small datasets, making it difficult to extrapolate their results beyond the datasets they employ.

Recently, deep learning (DL), which has achieved state-of-the-art performance in many fields, including image recognition, computer vision, biomedical image and signal classification, and time series prediction, has been adopted for arousal and sleep stage detection [38, 39, 40]. DL models learn to extract useful features directly from their inputs during training. As a result, they do not demand hand-crafted feature extraction, which is essential for classification accuracy and typically requires in-depth technical knowledge. Since the task is to assign a stage label to each epoch in the case of sleep stage scoring, researchers have employed classification algorithms including convolutional neural networks (CNN), recurrent neural networks (RNN), transformers, and hybrid networks with one or more epochs as inputs to determine the target epoch's stage. Chambon et al. [41] employed a CNN model containing 11 layers for sleep stage scoring from raw PSG signals. They used different numbers of epochs (from one to five) to detect the target epoch's stage. Khalili et al. [42] proposed a model with two CNNs, where the first one was used for extracting local features from each epoch and the second was employed to extract temporal features from the extracted feature vector. Phan et al. [43] introduced an RNN-based model accepting up to 30 epochs for sleep stage scoring. They implemented two RNNs, one for epoch modeling and the other for sequence modeling. Seo et al. [37] utilized a model with a CNN and an RNN, accepting multiple epochs and generating a stage label. Phan et al. [44] used a transformer model to detect the sleep stage of an epoch from 11 PSG epochs. Supratak et al. [45] proposed a two-step learning process for the detection of sleep stages. They used a CNN model to extract features from 25 epochs and two bidirectional long short-term memory (BiLSTM) layers, which is a type of RNN, for sequence modeling. Although these models achieved decent performances on variety of PSG datasets, they were developed to only perform sleep stage scoring tasks, and due to their architecture, they cannot be readily adapted for arousal detection.

In the case of arousal detection, scientists have introduced DL-based models benefiting from various preprocessing and ensemble techniques. Warric et al. [46] used the scattering transform and depthwise-separable convolution for feature extraction and BiLSTM layers for sequence. Pourbabaee et al. [47] introduced a model based on a dense convolutional neural network and BiLSTM. They denoised and normalized multi-channel PSG signals using an anti-aliasing filter and a sliding 18-minute window. The model also exploited auxiliary tasks to improve its detection performance. Liu et al. [48] employed multiple CNNs with preprocessed and segmented PSG signals to obtain initial classification and a random forest classifier using preliminary results for ensemble voting. Zhou et al. [49] downsampled and segmented continuous PSG signals before classifying them by a DL model consisting of CNNs and attention-based RNNs. Li et al. [50] proposed an approach similar to [49] but without the use of an attention mechanism. Li and Guan [51] developed a CNN model with different filter sizes and scales to learn both local and global interdependencies across an entire sleep record. They employ a novel augmentation method to improve scoring performance. Although arousal and stage scoring are clinically two similar and intertwined tasks and commonly performed together, researchers have always focused on only one of the tasks in their studies. Even though [46, 47] utilized sleep stages during model training, their main idea was to capture additional context from sleep stages to improve arousal detection accuracy. Thus, a new unified approach that can efficiently and reliably detect arousals and sleep stages is required.

Fully convolutional network (FCN) was proposed by Shelhamer et al. [52] for semantic segmentation, i.e., classifying each pixel in an image to a category. Since then, it has been adopted in other areas such as image denoising [53], medical image segmentation [54], and time series classification [55]. Such a model takes the input image and processes it at the pixel level, producing a mask that assigns a class label to each pixel of the input image and has the same size as the input image. Because of this ability, it can be used for the segmentation of PSG signals, recognizing regions of interest, namely arousals and stages, and potentially other sleep-related events such as apneas and hypopneas.

In this paper, we propose a novel multi-task learning approach for arousal and sleep stage detection using fully convolutional neural networks. An FCN model (named FullSleepNet) that accepts a full-night single-channel EEG signal and produces arousal and stage labels for the input was formed. FullSleepNet is made up of four modules: 1) a convolutional module to extract local features, 2) a recurrent module to construct long-range interdependencies, 3) an attention mechanism to focus on relevant parts of the input, and 4) a segmentation module to output final predictions. The model was simultaneously trained for arousal and sleep stage scoring tasks in an end-to-end manner. FullSleepNet was developed and evaluated using two large-scale, manually scored PSG datasets, i.e., the Sleep Heart Health Study [56, 57] and the Multi-Ethnic Study of Atherosclerosis [56, 58].

The summary of the study is as follows:

- To the best of our knowledge, this is the first study to propose a one-dimensional fully convolutional neural network model that unifies both arousal detection and sleep stage classification tasks. Furthermore, instead of treating both tasks

- as classification problems, we adopted a more suitable segmentation approach, particularly for arousal detection.
- Since output masks with the same size as the inputs are not required, we modified the FCN structure and discarded upsampling and deconvolution layers to reduce the use of computing resources. The model was designed to produce prediction masks with a resolution of 2 seconds.
- An end-to-end model was simultaneously trained for both tasks using only raw single-channel EEG signals. To minimize the complexity of preprocessing, a simple normalization was applied to signals.
- We evaluated FullSleepNet on two large-scale PSG datasets and provided epoch- and sample-level results for the arousal task. FullSleepNet achieved state-of-the-art performance in arousal detection, while the performance of the model in sleep stage classification was on par with existing methods.
- An ablation study was conducted to investigate the contribution of individual modules to the overall performance.

The rest of this paper is organized as follows. The datasets used in this study are detailed in Section 2. Section 3 describes the research methodology by elaborating on the multi-task learning strategy for sleep scoring that we present. Also, the FCN-based model structure is explained. Section 4 describes the experimental setup and gives a summary of the findings. Lastly, Sections 5 and 6 present the work's discussion and conclusion, respectively.

## 2. Datasets

Two large-scale datasets, i.e., the Sleep Heart Health Study (SHHS) [56, 57] and the Multi-Ethnic Study of Atherosclerosis (MESA) [56, 58], were employed to train and evaluate the arousal and sleep stage scoring performance of FullSleepNet. Table 1 and Table 2 list class distributions for sleep stages and arousals, respectively.

Table 1. The class-wise number of sleep stage epochs in the datasets

| **Dataset** | **W** | **N1** | **N2** | **N3** | **REM** |
|---|---|---|---|---|---|
| **SHHS** | 1,574,828 (28.4%) | 207,015 (3.7%) | 2,223,300 (40.1%) | 737,024 (13.3%) | 796,717 (14.4) |
| **MESA** | 431,663 (28.4%) | 135,599 (8.9%) | 633,198 (41.6%) | 123,704 (8.1%) | 197,186 (13.0%) |

Table 2. Sample- and epoch-level arousal distributions of the datasets

| **Dataset** | **Sample-level** | | **Epoch-level** | |
|---|---|---|---|---|
| | **Arousal** | **Non-arousal** | **Arousal** | **Non-arousal** |
| **SHHS** | 4.2% | 95.8% | 15.7% | 84.3% |
| **MESA** | 5.6% | 94.4% | 18.4% | 81.6% |

## 2.1. Sleep Heart Health Study (SHHS)

The SHHS dataset is a collection of data from different centers that studies how breathing problems during sleep affect heart diseases. The dataset has two sets of PSG recordings: Visit 1 (SHHS-1) with 5793 recordings and Visit 2 (SHHS-2) with 2651 recordings. Each recording has data from different sensors that measure brain waves, eye movements, muscle activity, heart rhythm, chest and abdomen movements, body position, light exposure, blood oxygen level, and airflow. Each epoch of recordings was manually annotated by sleep experts as W, N1, N2, N3, N4, REM, Movement, or Unknown as per Rechtschaffen and Kales (R&K) rules [59]. In parallel to sleep stage labels, arousal onset times and durations for each recording were determined by the experts. More information about the montage and scoring rules is described in [60]. In this study, we used single-channel EEG signals (C4-A1) with a sampling frequency of 125 Hz from SHHS-1. We excluded Movement and Unknown epochs and merged N3 and N4 based on AASM criteria, as in previous studies [37, 61]. Finally, we excluded a few recordings that lacked a complete set of AASM-defined sleep stages [61].

## 2.2. Multi-Ethnic Study of Atherosclerosis (MESA)

The MESA study is supported by the National Heart, Lung, and Blood Institute to track individuals from various centers over time and investigate factors associated with subclinical cardiovascular disease and the progression from subclinical to clinical disease [56, 58]. In the MESA Sleep Exam, 2056 participants underwent PSG at home using the Compumedics Somte System (Compumedics Ltd., Abbotsford, Australia) with different sensors to monitor the heart, brain, eye, chin, chest, abdomen, leg, snoring, and oxygen levels. Similar to the SHHS dataset, each recording was scored according to R&K rules, and arousal regions were marked by sleep specialists. More information about the montage and scoring rules is described in [62]. In this study, single-channel EEG data (C4-M1) sampled at 256 Hz was used. In line with previous studies [50], we excluded recordings that did not have at least five unique sleep stages or did not have arousal labels. Since the dataset contains some recordings that are unusually long, we used only thirty minutes before and after the sleep period for those. Finally, we downsampled EEG data by a factor of 2 to have a similar sampling frequency as the SHHS dataset.

## 3. Methodology

In this section, a brief discussion about the multi-task learning framework for arousal and sleep stage detection was done, the structure of the proposed model was introduced, and model training and evaluation processes were elaborated.

## 3.1. The framework

This paper presents a multi-task learning method for detecting arousals and sleep stages using FCNs. We design an FCN model that takes a single-channel EEG signal recorded during a whole night sleep as input and generates segmentation masks for arousals and sleep stages. Unlike many other studies that process each PSG recording epoch by epoch, FullSleepNet extracts discriminative local features from the full-length recording using convolution layers and constructs long-term dependencies among extracted features that represent the whole data with RNNs and an attention mechanism. Furthermore, instead of employing dense layers to classify extracted features for either arousals or sleep stages, the model produces predictions for both tasks using the segmentation module, which consists of two convolution layers, each followed by activations. Figure 1 illustrates the block diagram of the proposed approach.

Sleep experts visually analyze PSG signals to label 30-second-long epochs of sleep, starting from the first epoch and following rules defined by the AASM. To determine the sleep stage for an epoch, they look for the presence or absence of certain frequencies and predefined micro-events, such as sleep spindles, K-complexes, and delta brushes, in the epoch. Additionally, the AASM transition rules [12] dictate that both the preceding and succeeding epochs should be considered. Therefore, FullSleepNet employs a convolutional module to extract local features that correspond to sleep-related micro-events, as well as recurrent and attention modules to construct global features that represent inter-epoch context, mimicking the manual scoring of sleep experts.

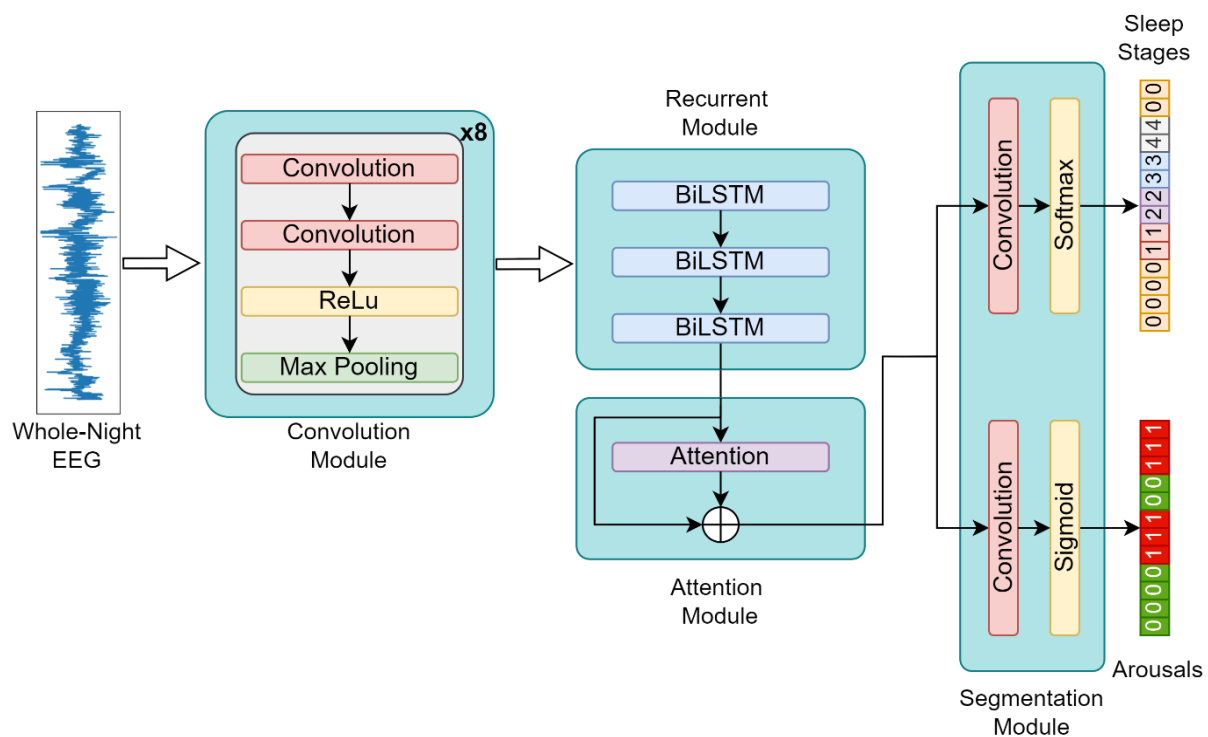

Figure 1. Block diagram of the proposed framework.

The AASM defines an arousal as a shift in EEG frequencies that includes alpha, theta, or frequencies greater than 16 Hz, lasts for at least three seconds, and is preceded by at least 10 seconds of continuous sleep. Unlike sleep stage scoring, which is based on epochs or fixed-length signal segments, arousal scoring involves determining the onset time and duration of each arousal that occurs during sleep. This requires locating arousal regions on the PSG, making it more similar to a segmentation problem than a classification problem. Since the sleep stage scoring task can be easily converted to a segmentation problem by upsampling ground truth stage labels, FullSleepNet unifies both tasks by utilizing the segmentation module that simultaneously produces predictions for both tasks.

Furthermore, AASM states that for an arousal to be scored during REM sleep, there must also be an increase in the submental EEG for at least one second. Therefore, arousal scoring depends not only on specific patterns and the presence of preceding sleep but also on sleep stages. Similarly, the occurrence of an arousal during one epoch influences the decision regarding the sleep stage of the subsequent epoch. Consequently, the proposed approach offers a unified solution for these technically different but clinically interrelated problems, as well as aims to improve the training process and overall performance through joint modeling and optimization.

### 3.2. Model Structure

FullSleepNet is specifically designed to process entire-night EEG signals and to produce prediction masks to segment signals based on arousals and sleep stages. The architecture comprises four primary modules, namely convolution, recurrent, attention, and segmentation modules, as illustrated in Figure 1. The convolution module, which consists of eight convolution blocks, extracts local features directly from the raw EEG signals to learn discriminative local features. Each block has two convolution layers with different kernel sizes, where a small filter captures temporal information and a large filter captures frequency information [45]. The extracted features are then passed through a ReLu activation function to introduce non-linearity and enable the model to learn complex representations of the input data. Additionally, a max-pooling layer is incorporated to reduce the dimensionality of extracted features and prevent overfitting, thus lowering the computational cost of the subsequent layers. In the subsequent blocks, the kernel size of convolution layers is reduced from (11, 9) to (5, 3) for faster training, and the number of filters is increased from (32, 32) to (192, 192) to extract more intricate features. As a result, a raw whole-night EEG $\mathbf{x} \in \mathbb{R}^{L \times 1}$ is represented by a feature map $\mathbf{f} \in \mathbb{R}^{L/256 \times 192}$ at the end of the convolution module, where $L$ is the zero-padded sequence length of the input EEG ($L = 2^{22}$ for SHHS and $2^{23}$ for MESA) and 256 is the downsampling factor for the temporal resolution, which results from the use of eight convolution blocks.

In the recurrent module, the model uses three successive BiLSTM layers with 128 units to learn long-range dependencies of extracted features in both forward and backward directions. For the input sequence $\mathbf{f} = \{\mathbf{f_1}, \mathbf{f_2}, \dots, \mathbf{f_T}\}$, the BiLSTM layer produces a sequence of hidden state vectors $\mathbf{h} = \{\mathbf{h_1}, \mathbf{h_2}, \dots, \mathbf{h_T}\}$ of the same length by iterating the following equations from $t = 1$ to $T$.

$$\vec{\mathbf{h}}_\mathbf{t}, \vec{\mathbf{c}}_\mathbf{t} = \overrightarrow{LSTM}(\vec{\mathbf{h}}_{\mathbf{t-1}}, \vec{\mathbf{c}}_{\mathbf{t-1}}, \mathbf{f_t}) \tag{1}$$

$$\overleftarrow{\mathbf{h}}_\mathbf{t}, \overleftarrow{\mathbf{c}}_\mathbf{t} = \overleftarrow{LSTM}(\overleftarrow{\mathbf{h}}_{\mathbf{t+1}}, \overleftarrow{\mathbf{c}}_{\mathbf{t+1}}, \mathbf{f_t}) \tag{2}$$

$$\mathbf{h_t} = \vec{\mathbf{h}}_\mathbf{t} || \overleftarrow{\mathbf{h}}_\mathbf{t} \tag{3}$$

where $T = L/256$ or $T = 2^{14}$ for SHHS and $2^{15}$ for MESA; $LSTM$ represents a function that processes the input vector $\mathbf{f_t}$ using a two-layer LSTM parameterized for forward and backward directions; $\vec{\mathbf{h}}_t$ and $\overleftarrow{\mathbf{h}}_t$ are the hidden state vectors of the BiLSTM in forward and backward directions, respectively; $\vec{\mathbf{c}}_t$ and $\overleftarrow{\mathbf{c}}_t$ are cell state vectors of the BiLSTM in forward and backward directions, respectively; $\vec{\mathbf{h}}_0, \vec{\mathbf{c}}_0, \overleftarrow{\mathbf{h}}_{T+1}$ and $\overleftarrow{\mathbf{c}}_{T+1}$ are zero vectors; $\vec{\mathbf{h}}_t || \overleftarrow{\mathbf{h}}_t$ denotes concatenation of hidden state vectors.

An attention module makes use of the attention mechanism, which is essentially a "neural network within a neural network" that mimics the cognitive attention of humans [63]. It enhances the important parts of its input and fades out the rest, so that the network can focus more on the relevant information and less on the noise [63]. In this work, we use the attention mechanism reported in [64]. The attention mechanism calculates the context vector in three steps. First, alignment scores are calculated as follows:

$$\mathbf{S_t} = \tanh(\mathbf{h_t} \cdot \mathbf{W_{att}} + \mathbf{b_{att}}) \qquad (4)$$

where $\mathbf{h_t}$ is the output of the recurrent module at time index $t$, and $\mathbf{W_{att}}$ and $\mathbf{b_{att}}$ are trainable weights and bias, respectively. After obtaining the $\mathbf{S_t}$ scores, the Softmax function is applied to compute attention weights $\mathbf{\alpha_t}$, which is calculated using the given formula:

$$\mathbf{\alpha_t} = \text{softmax}(\mathbf{S_t}) \qquad (5)$$

Then, the context vector $\mathbf{a}$ is calculated by the weighted sum of the hidden states as:

$$\mathbf{a} = \sum_{t=1}^{T} \mathbf{\alpha_t} \mathbf{h_t} \qquad (6)$$

The content vector $\mathbf{a}$ is summed with the output of the recurrent module $\mathbf{h}$ as $\tilde{\mathbf{h}} = \mathbf{h} + \mathbf{a}$ and fed to the segmentation module. The segmentation module consists of two branches, each containing a convolution layer followed by an activation function. Both convolution layers have a kernel size of 1, while the number of filters and type of activation function depend on the class counts of each problem. The first branch has one filter and generates a mask of probabilities $\hat{y}^a \in \mathbb{R}^{T \times 1}$ for arousals using the sigmoid activation function, where $T = 2^{14}$ for SHHS and $2^{15}$ for MESA. The second branch, on the other hand, has five filters, the same as the number of sleep stages, and generates another mask of class probabilities $\hat{y}^s \in \mathbb{R}^{T \times 5}$ for sleep stages using the softmax activation function.

### 3.3. Loss Calculation

To update weights and biases, i.e., trainable parameters of the model during training, a differentiable loss function, which measures the deviation of the predictions from ground truth labels, should be calculated. We use binary cross-entropy and categorical cross-entropy for arousals and sleep stages, respectively. The final loss is calculated as the weighted sum of both losses, as follows:

$$\mathcal{L} = w_1 \left[ -\frac{1}{N} \sum_{i=1}^{N} y_i^a \log \hat{y}_i^a + (1 - y_i^a) \log(1 - \hat{y}_i^a) \right] + w_2 \left[ -\sum_{i=1}^{N} y_i^s \log \hat{y}_i^s \right] \quad (7)$$

where $w_1$ and $w_2$ are weights, $y^a$ denotes ground truth arousal labels, and $y^s$ denotes one-hot encoded ground truth sleep stage labels. In this study, to give both tasks the same weight, we set $w_1 = w_2 = 1$. Note that, to compute the losses, it is necessary for the sizes of the ground truth label and prediction mask to be the same. Hence, the ground truth arousal labels were zero-padded and downsampled, while the ground truth sleep stage labels were upsampled and zero-padded.

### 3.4. Training

In line with previous literature [37, 51], the datasets were randomly split into training, validation, and test subsets using ratios of 0.5, 0.2, and 0.3, respectively. Each EEG signal was standardized by subtracting the mean from each data point and then dividing the result by the standard deviation. For SHHS data, each signal was zero-padded to the length of the nearest power of 2 to the length of the longest signal (i.e., $2^{22}$). For MESA data, each signal was downsampled by a factor of 2 to achieve a resolution similar to that of SHHS and then similarly zero-padded to a length of $2^{23}$.

The model was trained end-to-end using preprocessed data and resampled arousal and sleep stage labels. The Adam optimizer [65] was employed to optimize the loss, as defined in Section 3.3, while the learning rate was set to $10^{-4}$ and other parameters were set to their default values ($\beta_1 = 0.9$, $\beta_2 = 0.999$, and $\epsilon = 10^{-7}$). A simple data augmentation technique, which involved multiplying each signal by a random scalar in the range of [0.9, 1.1], was applied to training data to increase the model's generalization ability. To avoid overfitting, the training process was stopped early by monitoring the validation loss. Specifically, the training was terminated after 20 consecutive epochs with no improvement in the validation loss. The best-performing model in terms of validation loss was used to make predictions on the test sets. Python 3.8 and TensorFlow 2.5 [66] were used to implement the model and track the training progress.

### 3.5. Evaluation

As explained in Section 3.2, FullSleepNet produces two prediction masks with one-eighth the resolution of the input EEG, resulting in a shorter arousal mask with respect to the ground truth arousal labels and a longer sleep stage mask with respect to the ground truth sleep stage labels. Therefore, to accurately evaluate the performance of the model, the arousal prediction masks were upsampled, and the sleep stage prediction masks were downsampled to match the same length as their respective labels. As a result, performance metrics were calculated based on original-length ground truth labels and resampled prediction labels. Additionally, besides the sample-level calculation based on every sample point, we provided epoch-level metrics, which were calculated based on the presence or absence of arousals within 30-second epochs.

We used several metrics to evaluate the performance of our model, including Accuracy ($ACC$), F1 score ($F1$), and Cohen's kappa coefficient ($\kappa$) as epoch-level arousal and sleep stage metrics. In addition, we used area under the precision-recall curve ($AUPRC$) and area under the receiver operating characteristic curve ($AUROC$) as sample-level arousal metrics, as they are commonly employed metrics for this task [51].

Accuracy is defined as the ratio of correctly classified epochs to the total number of epochs in the dataset, and is formulated as:

$$ACC = \frac{TP + TN}{TP + FP + TN + FN} \tag{8}$$

where $TP, FP, TN$, and $FN$ represent the numbers of true positives, false positives, true negatives, and false negatives, respectively. While accuracy can be useful, it may not be the best measure of performance in datasets where one or more classes are significantly more prevalent than the others. In such cases, metrics like F1 score and Cohen's kappa coefficient may provide a more comprehensive evaluation of the model's performance.

F1 score is the harmonic mean of Precision ($PR$) and Recall ($RE$) and defined as follow:

$$F1 = 2\frac{PR \times RE}{PR + RE} \tag{9}$$

where

$$PR = \frac{TP}{TP + FP} \tag{10}$$

$$RE = \frac{TP}{TP + FN} \tag{11}$$

Cohen's kappa coefficient is a statistical measure of inter-rater agreement and calculated as the ratio of the observed agreement between the raters to the maximum possible agreement that could have occurred by chance:

$$\kappa = \frac{p_o - p_e}{1 - p_e} \tag{12}$$

where $p_o$ is observed agreement proportion between raters and $p_e$ is hypothetical chance agreement proportion between raters.

$AUPRC$ measures the trade-off between precision and recall for a binary classification problem and is calculated as the area under the precision-recall curve. It is particularly useful when the positive class is rare, and the goal is to accurately identify as many true positives as possible while minimizing false positives.

$AUROC$, on the other hand, measures the trade-off between true positive rate and false positive rate for a binary classification problem and is calculated as the area under the receiver operating characteristic curve. It is useful for evaluating a classifier's overall ability to discriminate between the positive and negative classes and is insensitive to class imbalance.

## 4. Results

In this study, an end-to-end deep learning model based on fully convolutional networks for the detection of arousals and sleep stages was presented. The model was trained using a multi-task learning approach and evaluated on two large-scale datasets. In this section, we present performance results for both tasks and visually demonstrate the model's predictions.

### 4.1. Results for Arousal Detection

In terms of sample-level metrics, FullSleepNet demonstrated high performance on the test sets of both SHHS and MESA datasets. Specifically, on SHHS dataset, FullSleepNet achieved an $AUPRC$ of 0.695 and an $AUROC$ of 0.973, while on MESA dataset, the $AUPRC$ and $AUROC$ were 0.696 and 0.962, respectively. Figure 2 illustrates the distribution of $AUPRC$ and $AUROC$ scores across records for both datasets. Despite the high variability in scores, only a small fraction of the test records had relatively low scores. For instance, 99% of the test records of SHHS had $AUPRC$ scores higher than 0.4, with 96% of them having scores above 0.5. Similarly, 96% of the test records for MESA had $AUPRC$ scores higher than 0.4, with 92% of them having scores above 0.5. Additionally, only eight test records from SHHS and 12 records from MESA had an $AUROC$ lower than 0.8, with 98% of SHHS and 94% of MESA test records having scores higher than 0.9.

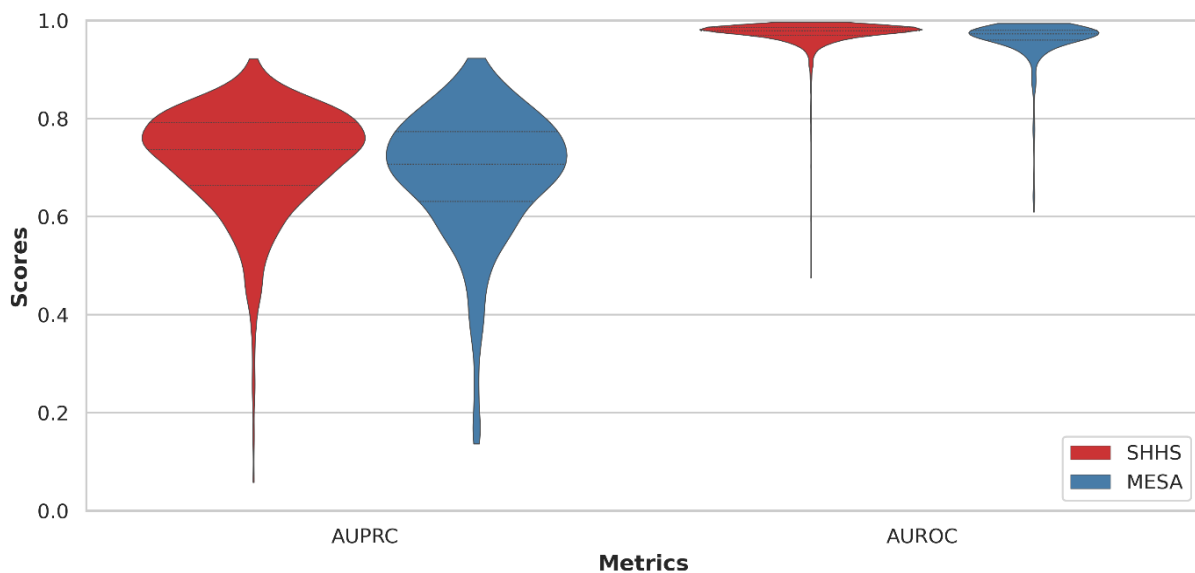

Figure 2. Record-wise distribution of $AUPRC$ and $AUROC$ for the datasets. The shapes of violin plots represent distributions of scores. The dashed lines in the plots correspond to the 25, 50, and 75 percentiles, respectively.

Table 3 presents FullSleepNet's epoch-level performance outcomes for both SHHS and MESA datasets, where the model's precision, recall, F1 score, accuracy, and kappa are listed. Notably, on both datasets, FullSleepNet achieved satisfactory performance in terms of all the evaluation metrics. Specifically, the model achieved a precision of 0.801, a recall of 0.683, an F1 score of 0.737, an accuracy of 0.923, and a kappa of 0.693 on SHHS dataset. Similarly, FullSleepNet obtained a precision of 0.803, a recall of 0.666, an F1 score of 0.728, an accuracy of 0.908, and a kappa of 0.674 on MESA dataset.

Table 3. Epoch-level performance results for SHHS and MESA dataset

| Dataset | PR | RE | F1 | ACC | κ |
| --- | --- | --- | --- | --- | --- |
| SHHS | 0.801 | 0.683 | 0.737 | 0.923 | 0.693 |
| MESA | 0.803 | 0.666 | 0.728 | 0.908 | 0.674 |

In Figure 3, we present two examples of FullSleepNet's arousal detection capabilities, showcasing its ability to detect both short and long arousal events. Specifically, Figure 3a displays a long arousal event lasting for more than 15 seconds, along with the corresponding output probabilities of FullSleepNet. Similarly, Figure 3b shows a short arousal event that lasts for less than six seconds. In both cases, the model was able to accurately capture the high-frequency EEG patterns associated with the events, demonstrating the effectiveness of FullSleepNet in detecting arousals of varying durations. It is worth noting that the output probabilities in both examples are continuous, gradually increasing at the beginning of the events, and then gradually decreasing as the events come to an end.

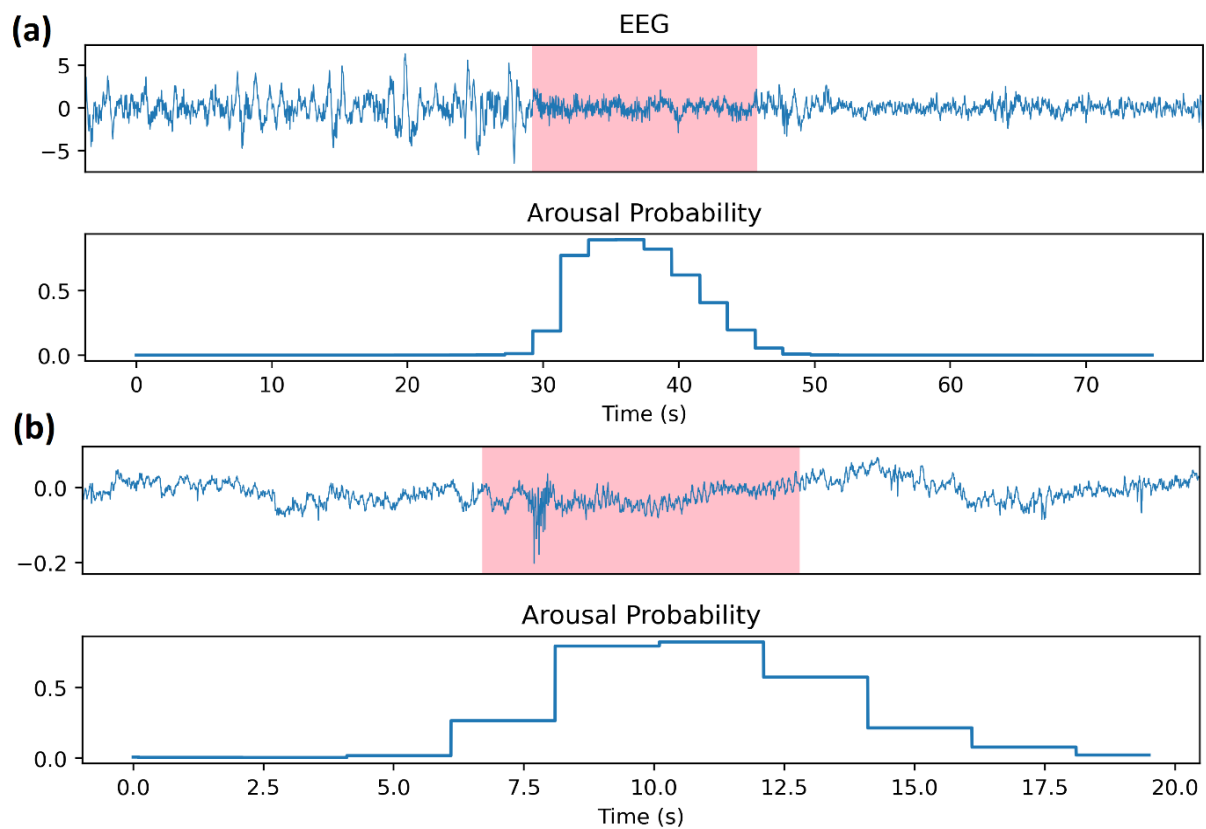

Figure 3. Demonstrative examples of the arousal detection. The pink shadows on the EEG curves indicate ground truth arousals. The arousal probability curves are upsampled arousal prediction masks of FullSleepNet. (a) a long arousal (>15 s) from SHHS data. (b) a short arousal (<6 s) from MESA data.

### 4.2. Results for Sleep Stage Detection

Figure 4 depicts confusion matrices for SHHS and MESA datasets, based on ground truth sleep stage labels and FullSleepNet's predictions. In SHHS dataset, Wake was the stage with the highest detection rate, correctly identified 93% of the time, while N1 was the most misclassified stage, with a detection rate of 43%. The classification rates for N2, N3, and REM were 89%, 81%, and 91%, respectively. Notably, N1 was frequently confused with N2 and Wake stages, while N3 was almost exclusively confused with N2. Similarly, in MESA dataset, Wake had the highest detection rate at 90%, while N1 had the lowest at 44%. The detection rates for N2, N3, and REM were 90%, 61%, and 86%, respectively. Although the model showed similar performance on both datasets, the detection rate of N3 for MESA was significantly lower than that of SHHS.

Figure 4. Confusion matrices for sleep stage detection on the (a) SHHS and (b) MESA datasets. The diagonal elements represent the number of correctly classified epochs, while the off-diagonal elements indicate the number of misclassified epochs. Color intensity reflects the percentage of correct predictions for each sleep stage, with darker shades indicating higher values.

Table 4. Sleep stage performance results of FullSleepNet on SHHS and MESA datasets

| Dataset | Stage | Precision | Recall | F1 Score |
|---|---|---|---|---|
| SHHS | W | 0.924 | 0.932 | 0.928 |
|  | N1 | 0.558 | 0.428 | 0.485 |
|  | N2 | 0.874 | 0.886 | 0.880 |
|  | N3 | 0.833 | 0.814 | 0.824 |
|  | REM | 0.886 | 0.911 | 0.898 |
| MESA | W | 0.900 | 0.896 | 0.898 |

|     |       |       |       |
| --- | ----- | ----- | ----- |
| N1  | 0.634 | 0.444 | 0.522 |
| N2  | 0.814 | 0.899 | 0.855 |
| N3  | 0.750 | 0.607 | 0.671 |
| REM | 0.862 | 0.860 | 0.861 |

Table 4 provides a detailed breakdown of FullSleepNet's performance on both SHHS and MESA datasets, including class-wise metrics such as precision, recall, and F1 scores for each sleep stage. In the case of SHHS dataset, the model obtained high precision and recall for Wake stage, with a precision of 0.924 and a recall of 0.932, resulting in an F1 score of 0.928. However, N1 stage had a lower precision of 0.558 and a recall of 0.428, leading to an F1 score of only 0.485. Similarly, for MESA dataset, the model demonstrated high precision and recall for Wake stage, with a precision of 0.900 and a recall of 0.896, resulting in an F1 score of 0.898. However, N1 stage had a lower precision of 0.634 and a recall of 0.444, leading to an F1 score of 0.522. In terms of overall metrics, accuracy, F1 score (macro-averaged), and Cohen's kappa coefficient were recorded for SHHS dataset as 0.875, 0.803, and 0.826, respectively. For MESA, FullSleepNet achieved an accuracy of 0.829, an F1 score of 0.761, and a kappa of 0.755. Finally, for the purpose of illustration, two examples comparing predictions with ground truth labels are presented in Figure 5, in which the model predictions are in good agreement with human expert's scores.

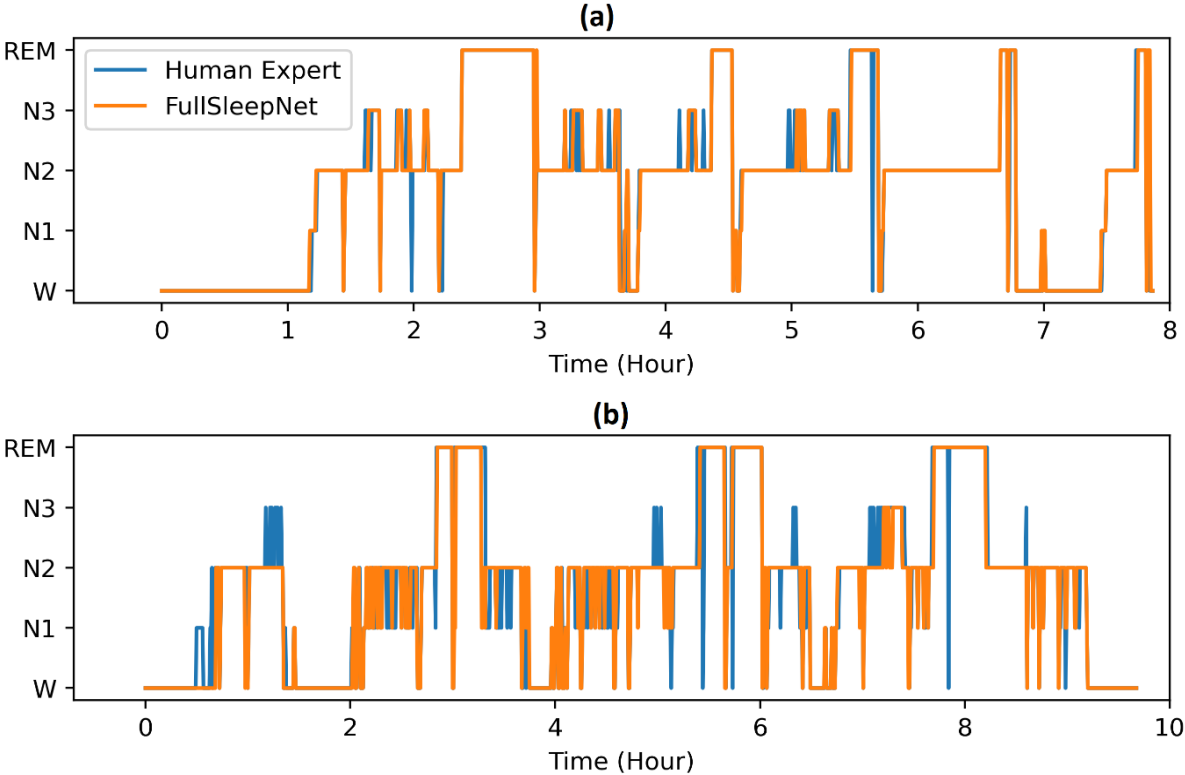

Figure 5. Hypnogram comparison between ground truth labels and predictions for two random participants from (a) SHHS and (b) MESA.

## 5. Discussion

This study presented an end-to-end deep learning model based on fully convolutional networks for the detection of arousals and sleep stages that was trained using a multi-task learning approach and evaluated on two large-scale datasets. Inspired by the semantic segmentation of images, our model was based on the segmentation of full-night single-channel EEG signals, unifying two tasks that are clinically interrelated but technically different as segmentation problems. The model demonstrated high performance on both tasks, with an $AUPRC$ of 0.695 and an $AUROC$ of 0.973 for the arousal detection on SHHS dataset and an $AUPRC$ of 0.696 and an $AUROC$ of 0.962 on MESA dataset. In addition, examples demonstrating the model's capabilities in detecting arousals of varying durations were provided. In terms of sleep stage detection performance, FullSleepNet achieved an accuracy of 0.875, an F1 score of 0.803, and a Cohen's kappa coefficient of 0.826. However, the model's performance was slightly lower on MESA dataset, with 0.829 accuracy, 0.761 F1 score, and 0.755 Cohen's kappa coefficient. Finally, confusion matrices were presented for both datasets to illustrate the model's performance on sleep stage detection.

FullSleepNet is a novel deep learning model for sleep analysis, capable of simultaneously performing arousal and sleep stage detection from raw EEG signals. Unlike traditional methods that rely on manual feature extraction and preprocessing techniques, FullSleepNet leverages convolution, recurrent, and attention modules to learn discriminative features directly from its input. To minimize the computational requirements of the model, we chose not to use extensive preprocessing methods, signal processing techniques, or signal transformations. Instead, we relied on features learned from standardized raw EEG signals. This not only reduces the computational demands of the model but also ensures that the extracted features are relevant to the problem at hand.

In contrast to other methods that perform each task separately, FullSleepNet takes a comprehensive approach by using the segmentation module to output two prediction masks: one for arousals and one for sleep stages. This module segments the input signal according to the different sleep events, providing a more detailed sleep analysis. Detecting both tasks at the same time also enables end-to-end training of the model, which can improve its performance by allowing it to learn relevant features for both tasks simultaneously.

Furthermore, FullSleepNet can process full-length signals, eliminating the need for sliding window methods that have been employed in other studies. By processing the entire signal at once, FullSleepNet is able to capture the temporal information present in the data and provide accurate predictions for the detection of arousals and sleep stages. This approach avoids the information loss that can occur at the boundaries of the windows used in sliding window methods.

Thanks to its efficient, data-driven, and comprehensive approach, our model demonstrated satisfactory performance for both tasks. Furthermore, the model demonstrated consistent and robust performance on two large-scale datasets for the arousal detection task, indicating its high adaptability and generalizability across different datasets and populations. However, the model's sleep stage classification performance on the MESA dataset was slightly lower than on the SHHS dataset, as measured by the F1 score (0.761 vs. 0.803). A closer examination of class-wise metrics revealed consistent results across all sleep stages except for N3. This can be attributed to the fact that the MESA dataset had a lower percentage of epochs labeled as N3 by sleep experts compared to other datasets such as SHHS and Sleep-EDF Expanded [67]. This difference in distribution may be due to high inter-scorer variability among experts who scored the MESA data, indicating potential inconsistencies or errors in scoring rules for N3.

In terms of the errors made by the model, we have observed that they primarily occur between stages that are adjacent in the sleep cycle. For instance, N3 is often mistaken for N2, but rarely for N1. Similarly, N1, which is known for having low inter-rater agreement [61], can be wrongly identified as Wake, N2, or REM since these stages may have patterns that resemble N1, but not N3. Additionally, we have found that REM is more frequently misidentified as N2 than Wake. One possible explanation for this is that both REM and Wake share the characteristic of eye movement, and there is minimal frontal activity related to eye movement that is detectable on the C4 derivations.

## 5.1. Ablation Study

To gain a better understanding of the role of each component in our proposed model, we conducted an ablation study where we systematically removed each module from the model and evaluated its impact on arousal detection and sleep stage classification performance. Our goal was to identify the components that significantly contribute to the model's overall performance and gain insights into the underlying mechanisms of the model.

In Table 5, we present the results of the ablation study, where C, R, and A represent the convolution, recurrent, and attention modules, respectively. The table shows the performance of each model variant on both datasets for arousal detection and sleep stage classification, as measured by $AUPRC$, $AUROC$, $ACC$, $F1$, and $\kappa$ values. We found that the inclusion of all three modules, i.e., convolution, recurrent, and attention (Model-CRA), led to the best performance on both datasets, with the highest $AUPRC$, $AUROC$, $ACC$, $F1$, and $\kappa$ values. On the other hand, Model-C, which only includes the convolution module, resulted in the lowest performance on both tasks.

Our findings suggest that including all modules leads to the best performance, with the recurrent module playing a crucial role, particularly for arousal detection. However, the contribution of the attention module to performance may be limited for two possible reasons: firstly, the spontaneous nature of arousals may make it difficult for the module to focus on specific parts of its input; secondly, the attention module may be redundant with the recurrent module, leading to limited additional benefits when used in combination. Overall, the ablation study highlights the importance of each module in the model and provides insights into the optimal architecture for achieving high performance in both tasks.

Table 5. Results for the ablation study. C, R, and A represent the convolution, recurrent, and attention modules, respectively.

| Model | Dataset | Arousal Detection | | Sleep Stage Classification | | |
|---|---|---|---|---|---|---|
| | | *AUPRC* | *AUROC* | *ACC* | *F1* | *κ* |
| Model-CRA | SHHS | 0.695 | 0.973 | 0.875 | 0.803 | 0.826 |
| Model-CR | SHHS | 0.694 | 0.973 | 0.871 | 0.800 | 0.820 |
| Model-CA | SHHS | 0.556 | 0.948 | 0.850 | 0.691 | 0.787 |
| Model-C | SHHS | 0.551 | 0.946 | 0.836 | 0.678 | 0.766 |
| Model-CRA | MESA | 0.696 | 0.962 | 0.829 | 0.761 | 0.755 |
| Model-CR | MESA | 0.689 | 0.960 | 0.828 | 0.754 | 0.753 |
| Model-CA | MESA | 0.514 | 0.918 | 0.782 | 0.653 | 0.685 |
| Model-C | MESA | 0.506 | 0.912 | 0.776 | 0.656 | 0.674 |

## 5.2. Comparison With State-of-the-Art Approaches

Table 6 presents a comparison of FullSleepNet's arousal detection performance with state-of-the-art models, including their architectures, datasets, and PSG signals used for development. DeepCAD [50] is a deep learning model consisting of 33 convolutional layers, 2 long short-term memory (LSTM) layers, and a fully connected layer. The model was trained and tested on both SHHS and MESA datasets using single-lead, 1-hour-long ECG signals, with reported $AUPRC$ scores of 0.54 and 0.92, respectively. Although the use of ECG signals may be considered a disadvantage compared to EEGs, it's worth noting that their evaluation was based on labels that were trimmed to a range of 30 seconds before the first and 30 seconds after the last arousals and resampled to 1 Hz, which might have introduced some bias into their results.

DeepSleep [51] is a CNN-based model with 35 convolution layers. It accepts full-length multi-channel PSG signals, including EEG, EOG, ECG, EMG, SaO2, and airflow. Researchers trained three models, each accepting PSG signals with a different resolution. The ensemble of these three models was evaluated on 1000 recordings from SHHS dataset, with an $AURPC$ of 0.59 and an $AUROC$ of 0.94. Zhang et al. [68] proposed a CNN model similar to DeepSleep but with more layers, using various signals such as SaO2, ECG, heart rate, airflow, position, light, and oxygen saturation. The model was trained and evaluated on 1000 recordings from SHHS dataset, achieving an $AURPC$ of 0.56 and an $AUROC$ of 0.93. However, when they evaluated the model on 2-channel EEG signals, the model performed poorly.

In addition, Alvarez-Estevez and Fernández-Varela [69] introduced a method based on manual feature extraction, thresholding, and post-processing using EEG and EMG signals. They evaluated their algorithm on various datasets, including SHHS, and reported epoch-level F1 score of 0.610 and epoch-level Cohen's kappa coefficient of 0.573. Finally, we excluded studies that utilized only a small portion of the SHHS from the comparison. For instance, Álvarez-Estévez and Moret-Bonillo [25] employed 20 recordings, Álvarez-Estévez [29] used 10 recordings, and Ugur and Erdamar [30] worked with 1200 epochs of 5 subjects using manual feature extraction and classical machine learning algorithms.

Overall, FullSleepNet is a powerful and effective model for arousal detection in sleep studies. Its efficient, data-driven approach and comprehensive architecture enable it to adapt well to various datasets and generalize to new situations. The robustness of FullSleepNet is evident from the results of our experiments, which demonstrate that it significantly outperformed existing state-of-the-art models on both the SHHS and MESA datasets.

Table 6. Performance comparison between FullSleepNet and the state-of-the-art methods for arousal detection.

| Study | Architecture | Dataset | PSG Signal(s) | *AURPRC* | *AUROC* |
|---|---|---|---|---|---|
| DeepCAD [50] | CNN+RNN | SHHS | ECG | 0.54 | 0.92 |
| DeepSleep [51] | CNN | SHHS | EEG, EOG, ECG, EMG, SaO2, and Airflow | 0.59 | 0.94 |
| Zhang et al. [68] | CNN | SHHS | SaO2, ECG, Heart rate, Airflow, Position, Light, Saturation | 0.56 | 0.93 |
| Zhang et al. [68] | CNN | SHHS | EEG | 0.05 | 0.49 |
| **FullSleepNet** | CNN+RNN+FCN | SHHS | EEG | **0.70** | **0.97** |
| DeepCAD [50] | CNN+RNN | MESA | ECG | 0.62 | 0.93 |
| **FullSleepNet** | CNN+RNN+FCN | MESA | EEG | **0.70** | **0.96** |

Specifically, FullSleepNet improved the current best AUPRC values by 17% and 13% for SHHS and MESA datasets, respectively, using single-channel EEG signals. This demonstrates the model's ability to effectively identify sleep arousal events with high precision and recall. Moreover, the use of a single-channel EEG signal is a significant advantage of the FullSleepNet model, making it more practical and cost-effective for real-world applications.

Table 7 presents a comparison of FullSleepNet's sleep stage detection performance on SHHS dataset with state-of-the-art models. The table lists the model architectures, deep learning model input types, PSG signals, and the number of epochs used to score the target epoch's sleep stage. FullSleepNet exhibited slightly lower sleep stage performance than the best-performing models in terms of accuracy, F1 score, and kappa, with differences of only 0.2%, 0.5, and 0.2%, respectively, despite their use of short-time Fourier transform to generate spectrograms of EEG signals. Nevertheless, our model outperformed the model proposed by Zhang et al. [68], which is the only approach that handles both arousal and sleep stage detection tasks simultaneously, by a large margin, recording 17%, 20%, and 25% improvements in terms of accuracy, F1 score, and kappa, respectively. Overall, FullSleepNet exhibited comparable performance to other state-of-the-art methods while handling two tasks at the same time and using raw EEG signals as input.

Table 7. Performance comparison between FullSleepNet and the state-of-the-art methods for sleep scoring on SHHS dataset via deep learning.

| Model | Architecture | PSG Signal | Input Type | # of Epochs | ACC | F1 | κ |
|---|---|---|---|---|---|---|---|
| IITNet [37] | CNN+RNN | EEG | Raw | 4 | 0.867 | 0.798 | 0.810 |
| Sors et al. [61] | CNN | EEG | Raw | 4 | 0.868 | 0.785 | 0.810 |
| XSleepNet2 [70] | CNN+RNN | EEG | Raw+Spectrogram | 20 | 0.876 | **0.807** | 0.826 |
| SleepTransformer [44] | Transformer | EEG | Spectrogram | 11 | **0.877** | 0.801 | **0.828** |
| AttnSleep [71] | CNN+Attention | EEG | Raw | 3 | 0.866 | 0.797 | 0.810 |
| SeqSleepNet [72] | RNN+Attention | EEG | Spectrogram | 30 | 0.865 | 0.785 | 0.811 |
| Zhang et al. [68] | CNN | Various | Raw | All | 0.724 | 0.646 | 0.623 |
| **FullSleepNet** | CNN+RNN+FCN | EEG | Raw | All | 0.875 | 0.803 | 0.826 |

In summary, our study presented FullSleepNet, an end-to-end deep learning model for arousal detection and sleep stage classification. Our model achieved state-of-the-art performance for arousal detection and comparable performance for sleep stage classification on two large-scale datasets. Moreover, FullSleepNet offers several advantages over existing state-of-the-art models, which can be summarized as follows:

- FullSleepNet employs raw EEG signals as input, avoiding any time-consuming and subjective manual feature extraction and computationally expensive signal transformations.
- It unifies two clinically interrelated but technically different tasks as segmentation problems, making the model more practical and easier to use.
- It utilizes a multi-task learning approach, allowing the model to learn from the shared features of both tasks and improve overall performance.
- It does not require sliding window techniques, which reduces the computational cost and enables predictions for both tasks with a single forward pass. It can score a whole-night EEG signal on a modest PC with an AMD Ryzen 5 5500 3.70 GHz CPU, 16 GB of RAM, and no GPU in just 18 seconds.
- It processes full-length EEG signals, which preserves the temporal information of the signals and allows for better detection of transient events.

Overall, FullSleepNet's ability to handle both tasks, employ raw EEG signals, process full-length signals, and utilize a multi-task learning approach makes it a promising tool for accurate and efficient arousal detection and sleep stage classification. In the future, FullSleepNet can evolve into a full-fledged clinical decision support system for sleep medicine by incorporating the ability to score respiratory events, movements, and cardiac events.

## 6. Conclusion

In conclusion, FullSleepNet is an end-to-end deep learning model that offers improved practicality, efficiency, and accuracy for the detection of arousal and the classification of sleep stages. The model comprises four modules, including convolution, recurrent, attention, and segmentation, and uses raw EEG signals as input. By unifying the two interrelated tasks as segmentation problems and employing a multi-task learning approach, FullSleepNet achieves state-of-the-art performance for arousal detection and comparable performance for sleep stage classification on two large-scale datasets. Notably, FullSleepNet eliminates the need for sliding windows, reducing computational costs and improving prediction accuracy. Future research could explore FullSleepNet's potential for developing into a clinical decision support system for sleep medicine, incorporating the ability to score respiratory events, movements, and cardiac events.

## Data availability statement

At the time of publication, all code required for performing the above-described analysis will be made available in a GitHub repository https://github.com/hasanzan/FullSleepNet. The datasets used in this study can be accessed after permission is granted by the National Sleep Research Resource.

## Conflict of interest

All authors declare that they have no conflicts of interests.

## Acknowledgement

The numerical calculations reported in this paper were fully performed at TUBITAK ULAKBIM, High Performance and Grid Computing Center (TRUBA resources).

## References

[1] S. Mukherjee, S. R. Patel, S. N. Kales, N. T. Ayas, K. P. Strohl, D. Gozal and A. Malhotra, "An Official American Thoracic Society Statement: The Importance of Healthy Sleep. Recommendations and Future Priorities," *American Journal of Respiratory and Critical Care Medicine,* vol. 191, no. 12, pp. 1450-1458, 2018.

[2] B. B. Kamdar, D. M. Needham and N. A. Collop, "Sleep Deprivation in Critical Illness," *Journal of Intensive Care Medicine,* vol. 27, no. 2, pp. 97-111, 2012.

[3] P. McNamara, P. Johnson, D. McLaren, E. Harris, C. Beauharnais and S. Auerbach, "Rem And Nrem Sleep Mentation," *International Review of Neurobiology,* vol. 92, pp. 69-86, 2010.

[4] F. Guidozzi, "Sleep and sleep disorders in menopausal women," *Climacteric,* vol. 16, no. 2, pp. 214-219, 2013.

[5] P. Halasz, M. Terzano, L. Parrino and R. Bodizs, "The nature of arousal in sleep," *Journal of Sleep Research,* vol. 13, no. 1, pp. 1-23, 2004.

[6] M. Boselli, L. Parrino, A. Smerieri and M. G. Terzano, "Effect of Age on EEG Arousals in Normal Sleep," *Sleep,* vol. 21, no. 4, pp. 361-367, 1998.

[7] M. J. Thorpy, "Classification of Sleep Disorders," *Journal of Clinical Neurophysiology,* vol. 7, no. 1, pp. 67-82, 1990.

[8] L. M. McCracken and G. L. Iverson, "Disrupted Sleep Patterns and Daily Functioning in Patients with Chronic Pain," *Pain Research and Management,* vol. 7, no. 2, pp. 75-79, 2002.

[9] E. Hita-Yanez, M. Atienza, E. Gil-Neciga and J. L. Cantero, "Disturbed Sleep Patterns in Elders with Mild Cognitive Impairment: The Role of Memory Decline and ApoE ε4 Genotype," *Current Alzheimer Research,* vol. 9, no. 3, pp. 290-297, 2012.

[10] S. K. Howard, "Sleep Deprivation and Physician Performance: Why Should I Care?," *Baylor University Medical Center Proceedings,* vol. 18, no. 2, pp. 108-112, 2005.

[11] S. Lüdtke, W. Hermann, T. Kirste, H. Beneš and S. Teipel, "An algorithm for actigraphy-based sleep/wake scoring: Comparison with polysomnography," *Clinical Neurophysiology,* vol. 132, no. 1, pp. 137-145, 2021.

[12] H. Zan and A. Yildiz, "Local Pattern Transformation-Based convolutional neural network for sleep stage scoring," *Biomedical Signal Processing and Control,* vol. 80, p. 104275, 2023.

[13] L. Fiorillo, A. Puiatti, M. Papandrea, P.-L. Ratti, P. Favaro, C. Roth, P. Bargiotas, C. L. Bassetti and F. D. Faraci, "Automated sleep scoring: A review of the latest approaches," *Sleep Medicine Reviews,* vol. 48, p. 101204, 2019.

[14] M. Younes, S. T. Kuna, A. I. Pack, J. K. Walsh, C. A. Kushida, B. Staley and G. W. Pien, "Reliability of the American Academy of Sleep Medicine Rules for Assessing Sleep Depth in Clinical Practice," *Journal of Clinical Sleep Medicine,* vol. 14, no. 02, pp. 205-2013, 2018.

[15] X. Qian, Y. Qiu, Q. He, Y. Lu, H. Lin, F. Xu, F. Zhu, Z. Liu, X. Li, Y. Cao and J. Shuai, "A Review of Methods for Sleep Arousal Detection Using Polysomnographic Signals," *Brain Sciences,* vol. 11, no. 10, p. 1274, 2021.

[16] K. Aboalayon, M. Faezipour, W. Almuhammadi and S. Moslehpour, "Sleep Stage Classification Using EEG Signal Analysis: A Comprehensive Survey and New Investigation," *Entropy,* vol. 18, no. 9, p. 272, 2016.

[17] B. Koley and D. Dey, "An ensemble system for automatic sleep stage classification using single channel EEG signal," *Computers in Biology and Medicine,* vol. 42, no. 12, pp. 1186-1195, 2012.

[18] P. Memar and F. Faradji, "A Novel Multi-Class EEG-Based Sleep Stage Classification System," *IEEE Transactions on Neural Systems and Rehabilitation Engineering,* vol. 26, no. 1, pp. 84-95, 2018.

[19] A. R. Hassan and A. Subasi, "A decision support system for automated identification of sleep stages from single-channel EEG signals," *Knowledge-Based Systems,* vol. 128, pp. 115-124, 2017.

[20] Ş. Yücelbaş, C. Yücelbaş, G. Tezel, S. Özşen and Ş. Yosunkaya, "Automatic sleep staging based on SVD, VMD, HHT and morphological features of single-lead ECG signal," *Expert Systems with Applications,* vol. 102, pp. 193-206, 2018.

[21] M. Xiao, H. Yan, J. Song, Y. Yang and X. Yang, "Sleep stages classification based on heart rate variability and random forest," *Biomedical Signal Processing and Control,* vol. 8, no. 6, pp. 624-633, 2013.

[22] F. de Carli, L. Nobili, P. Gelcich and F. Ferrillo, "A Method for the Automatic Detection of Arousals During Sleep," *Sleep,* vol. 22, no. 5, pp. 561-572, 1999.

[23] G. Pillar, A. Bar, M. Betito, R. P. Schnall, I. Dvir, J. Sheffy and P. Lavie, "An automatic ambulatory device for detection of AASM defined arousals from sleep: the WP100," *Sleep Medicine,* vol. 4, no. 3, pp. 207-212, 2003.

[24] T. Sugi, F. Kawana and M. Nakamura, "Automatic EEG arousal detection for sleep apnea syndrome," *Biomedical Signal Processing and Control,* vol. 4, no. 4, pp. 329-337, 2009.

[25] D. Álvarez-Estévez and V. Moret-Bonillo, "Identification of Electroencephalographic Arousals in Multichannel Sleep Recordings," *IEEE Transactions on Biomedical Engineering,* vol. 58, no. 1, p. 54*63, 2011.

[26] P. Huy, D. Quan, D. The-Luan and V. Duc-Lung, "Metric learning for automatic sleep stage classification," in *2013 35th Annual International Conference of the IEEE Engineering in Medicine and Biology Society (EMBC)*, Osaka, Japan, 2013.

[27] I. Fernández-Varela, E. Hernández-Pereira, D. Álvarez-Estévez and V. Moret-Bonillo, "Combining machine learning models for the automatic detection of EEG arousals," *Neurocomputing,* vol. 268, pp. 100-108, 2017.

[28] T. Lajnef, S. Chaibi, P. Ruby, P.-E. Aguera, J.-B. Eichenlaub, M. Samet, A. Kachouri and K. Jerbi, "Learning machines and sleeping brains: Automatic sleep

stage classification using decision-tree multi-class support vector machines," *Journal of Neuroscience Methods,* vol. 250, pp. 94-105, 2015.

[29] D. Álvarez-Estévez, N. Sánchez-Maroño, A. Alonso-Betanzos and V. Moret-Bonillo, "Reducing dimensionality in a database of sleep EEG arousals," *Expert Systems with Applications,* vol. 38, no. 6, pp. 7746-7754, 2011.

[30] T. K. Ugur and A. Erdamar, "An efficient automatic arousals detection algorithm in single channel EEG," *Computer Methods and Programs in Biomedicine,* vol. 173, pp. 131-138, 2019.

[31] Z. Huang and B. W.-K. Ling, "Sleeping stage classification based on joint quaternion valued singular spectrum analysis and ensemble empirical mode decomposition," *Biomedical Signal Processing and Control,* vol. 71, p. 103086, 2022.

[32] A. R. Hassan and M. I. H. Bhuiyan, "Computer-aided sleep staging using Complete Ensemble Empirical Mode Decomposition with Adaptive Noise and bootstrap aggregating," *Biomedical Signal Processing and Control,* vol. 24, pp. 1-10, 2016.

[33] S. K. Satapathy, A. K. Bhoi, D. Loganathan, B. Khandelwal and P. Barsocchi, "Machine learning with ensemble stacking model for automated sleep staging using dual-channel EEG signal," *Biomedical Signal Processing and Control,* vol. 69, p. 102898, 2021.

[34] I. Koprinska, G. Pfurtscheller and D. Flotzinger, "Sleep classification in infants by decision tree-based neural networks," *Artificial Intelligence in Medicine,* vol. 8, no. 4, pp. 387-401, 1996.

[35] N. Schaltenbrand, R. Lengelle, M. Toussaint, R. Luthringer, G. Carelli, A. Jacqmin, E. Lainey, A. Muzet and J. P. Macher, "Sleep Stage Scoring Using the Neural Network Model: Comparison Between Visual and Automatic Analysis in Normal Subjects and Patients," *Sleep,* vol. 19, no. 1, pp. 26-35, 1 1996.

[36] I. Fernández-Varela, D. Alvarez-Estevez, E. Hernández-Pereira and V. Moret-Bonillo, "A simple and robust method for the automatic scoring of EEG arousals in polysomnographic recordings," *Computers in Biology and Medicine,* vol. 87, pp. 77-86, 2017.

[37] H. Seo, S. Back, S. Lee, D. Park, T. Kim and K. Lee, "Intra- and inter-epoch temporal context network (IITNet) using sub-epoch features for automatic sleep scoring on raw single-channel EEG," *Biomedical Signal Processing and Control,* vol. 61, p. 102037, 2020.

[38] K. Muhammad, J. Ahmad, Z. Lv, P. Bellavista, P. Yang and S. W. Baik, "Efficient Deep CNN-Based Fire Detection and Localization in Video Surveillance

Applications," *IEEE Transactions on Systems, Man, and Cybernetics: Systems,* vol. 49, no. 7, pp. 1419-1434, 2019.

[39] A. Ullah, J. Ahmad, K. Muhammad, M. Sajjad and S. W. Baik, "Action Recognition in Video Sequences using Deep Bi-Directional LSTM With CNN Features," *IEEE Access,* vol. 6, pp. 1155-1166, 2018.

[40] M. Sajjad, S. Khan, T. Hussain, K. Muhammad, A. K. Sangaiah, A. Castiglione, C. Esposito and S. W. Baik, "CNN-based anti-spoofing two-tier multi-factor authentication system," *Pattern Recognition Letters,* vol. 126, pp. 123-131, 2019.

[41] S. Chambon, M. N. Galtier, P. J. Arnal, G. Wainrib and A. Gramfort, "A Deep Learning Architecture for Temporal Sleep Stage Classification Using Multivariate and Multimodal Time Series," *IEEE Transactions on Neural Systems and Rehabilitation Engineering,* vol. 26, no. 4, pp. 758-769, 2018.

[42] E. Khalili and B. Mohammadzadeh Asl, "Automatic Sleep Stage Classification Using Temporal Convolutional Neural Network and New Data Augmentation Technique from Raw Single-Channel EEG," *Computer Methods and Programs in Biomedicine,* vol. 204, p. 106063, 2021.

[43] H. Phan, F. Andreotti, N. Cooray, O. Y. Chen and M. De Vos, "Joint Classification and Prediction CNN Framework for Automatic Sleep Stage Classification," *IEEE Transactions on Biomedical Engineering,* vol. 66, no. 5, pp. 1285-1296, 2019.

[44] H. Phan, K. Mikkelsen, O. Y. Chen, P. Koch, A. Mertins and M. De Vos, "SleepTransformer: Automatic Sleep Staging With Interpretability and Uncertainty Quantification," *IEEE Transactions on Biomedical Engineering,* vol. 69, no. 8, pp. 2456-2467, 2022.

[45] A. Supratak, H. Dong, C. Wu and Y. Guo, "DeepSleepNet: A Model for Automatic Sleep Stage Scoring Based on Raw Single-Channel EEG.," *IEEE transactions on neural systems and rehabilitation engineering : a publication of the IEEE Engineering in Medicine and Biology Society,* vol. 25, no. 11, pp. 1998-2008, 2017.

[46] P. A. Warrick, V. Lostanlen and M. Nabhan Homsi, "Hybrid scattering-LSTM networks for automated detection of sleep arousals," *Physiological Measurement,* vol. 40, no. 7, p. 074001, 2019.

[47] B. Pourbabaee, M. H. Patterson, M. R. Patterson and F. Benard, "SleepNet: automated sleep analysis via dense convolutional neural network using physiological time series," *Physiological Measurement,* vol. 40, no. 8, p. 084005, 2019.

[48] Y. Liu, H. Liu and B. Yang, "Automatic Sleep Arousals Detection From Polysomnography Using Multi-Convolution Neural Network and Random Forest," *IEEE Access,* vol. 8, pp. 176343-176350, 2020.

[49] G. Zhou, R. Li, S. Zhang, J. Wang and J. Ma, "Multimodal Sleep Signals-Based Automated Sleep Arousal Detection," *IEEE Access,* vol. 8, pp. 106157-106164, 2020.

[50] A. Li, S. Chen, S. F. Quan, L. S. Powers and J. M. Roveda, "A deep learning-based algorithm for detection of cortical arousal during sleep," *Sleep,* vol. 43, no. 12, 2020.

[51] H. Li and Y. Guan, "DeepSleep convolutional neural network allows accurate and fast detection of sleep arousal," *Communications Biology,* vol. 4, no. 1, p. 18, 2021.

[52] E. Shelhamer, J. Long and T. Darrell, "Fully Convolutional Networks for Semantic Segmentation," *IEEE Transactions on Pattern Analysis and Machine Intelligence,* vol. 39, no. 4, pp. 640-651, 2017.

[53] S.-F. Wang, W.-K. Yu and Y.-X. Li, "Multi-Wavelet Residual Dense Convolutional Neural Network for Image Denoising," *IEEE Access,* vol. 8, pp. 214413-214424, 2020.

[54] J. Dolz, C. Desrosiers and I. Ben Ayed, "3D fully convolutional networks for subcortical segmentation in MRI: A large-scale study," *NeuroImage,* vol. 170, pp. 456-470, 2018.

[55] F. Karim, S. Majumdar, H. Darabi and S. Chen, "LSTM Fully Convolutional Networks for Time Series Classification," *IEEE Access,* pp. 1662-1669, 2018.

[56] G.-Q. Zhang, L. Cui, R. Mueller, S. Tao, M. Kim, M. Rueschman, S. Mariani, D. Mobley and S. Redline, "The National Sleep Research Resource: towards a sleep data commons," *Journal of the American Medical Informatics Association,* vol. 25, no. 10, pp. 1351-1358, 2018.

[57] S. F. Quan, B. V. Howard, C. Iber, J. P. Kiley, F. J. Nieto, G. T. O'Connor, D. M. Rapoport, S. Redline, J. Robbins, J. M. Samet and P. W. Wahl, "The Sleep Heart Health Study: Design, Rationale, and Methods," *Sleep,* vol. 20, no. 12, pp. 1077-1085, 1997.

[58] X. Chen, R. Wang, P. Zee, P. L. Lutsey, S. Javaheri, C. Alcántara, C. L. Jackson, M. A. Williams and S. Redline, "Racial/Ethnic Differences in Sleep Disturbances: The Multi-Ethnic Study of Atherosclerosis (MESA)," *Sleep,* vol. 38, no. 6, pp. 877-888, 2015.

[59] E. A. Wolpert, "A Manual of Standardized Terminology, Techniques and Scoring System for Sleep Stages of Human Subjects.," *Archives of General Psychiatry,* vol. 20, no. 2, pp. 246-247, 1969.

[60] National Sleep Research Resource, "National Sleep Research Resource," [Online]. Available: https://sleepdata.org/datasets/shhs. [Accessed 14 March 2023].

[61] A. Sors, S. Bonnet, S. Mirek, L. Vercueil and J.-F. Payen, "A convolutional neural network for sleep stage scoring from raw single-channel EEG," *Biomedical Signal Processing and Control,* vol. 42, pp. 107-114, 2018.

[62] National Sleep Research Resource, "National Sleep Research Resource," [Online]. Available: https://sleepdata.org/datasets/mesa. [Accessed 15 March 2023].

[63] D. Soydaner, "Attention mechanism in neural networks: where it comes and where it goes," *Neural Computing and Applications,* vol. 34, no. 16, pp. 13371-13385, 2022.

[64] S. Hamdi, M. Oussalah, A. Moussaoui and M. Saidi, "Attention-based hybrid CNN-LSTM and spectral data augmentation for COVID-19 diagnosis from cough sound," *Journal of Intelligent Information Systems,* vol. 59, no. 2, pp. 367-389, 2022.

[65] D. P. Kingma and J. Ba, "Adam: A Method for Stochastic Optimization," *arXiv,* vol. 1412.6980, 2014.

[66] M. Abadi, A. Agarwal, P. Barham, E. Brevdo and Z. Chen, "TensorFlow: Large-Scale Machine Learning on Heterogeneous Distributed Systems," *arXiv Preprint,* vol. 1603.04467, 2016.

[67] A. L. Goldberger, L. A. N. Amaral, L. Glass, J. M. Hausdorff, P. C. Ivanov, R. G. Mark, J. E. Mietus, G. B. Moody, C.-K. Peng and H. E. Stanley, "PhysioBank, PhysioToolkit, and PhysioNet," *Circulation,* vol. 101, no. 23, 2000.

[68] H. Zhang, X. Wang, H. Li, S. Mehendale and Y. Guan, "Auto-annotating sleep stages based on polysomnographic data," *Patterns,* vol. 3, no. 1, p. 100371, 2022.

[69] D. Alvarez-Estevez and I. Fernández-Varela, "Large-scale validation of an automatic EEG arousal detection algorithm using different heterogeneous databases," *Sleep Medicine,* vol. 57, pp. 6-14, 2019.

[70] H. Phan, O. Y. Chen, M. C. Tran, P. Koch, A. Mertins and M. De Vos, "XSleepNet: Multi-View Sequential Model for Automatic Sleep Staging," *IEEE Transactions on Pattern Analysis and Machine Intelligence,* vol. 44, no. 9, pp. 5903-5915, 2011.

[71] E. Eldele, Z. Chen, C. Liu, M. Wu, C.-K. Kwoh, X. Li and C. Guan, "An Attention-Based Deep Learning Approach for Sleep Stage Classification With Single-Channel EEG," *IEEE Transactions on Neural Systems and Rehabilitation Engineering,* vol. 29, pp. 809-818, 2021.

[72] H. Phan, F. Andreotti, N. Cooray, O. Y. Chen and M. De Vos, "SeqSleepNet: End-to-End Hierarchical Recurrent Neural Network for Sequence-to-Sequence Automatic Sleep Staging," *IEEE Transactions on Neural Systems and Rehabilitation Engineering,* vol. 27, no. 3, pp. 400-410, 2019.